\documentstyle[11pt,moriond,epsf]{article}
\newcommand{\fig}[3]{\epsfxsize=#1\epsfysize=#2\epsfbox{#3}}

\def\Journal#1#2#3#4{{#1} {\bf #2}, #3 (#4)}

\def\NPB{{\em Nucl. Phys.} B}
\def\PLB{{\em Phys. Lett.}  B}

\def\be{\begin{equation}}
\def\ee{\end{equation}}
\def\bea{\begin{eqnarray}}
\def\eea{\end{eqnarray}}

\begin{document}
\begin{flushright}
LPTENS-98/23 \\
{\tt hep-ph@xxx/9805469}\\
May 1998
\end{flushright}

\vspace*{2cm}
\title{DUALITY, SUPERSTRINGS \& M-THEORY}

\author{A. BILAL}

\address{CNRS - Laboratoire de Physique Th\'eorique de l'\'Ecole
Normale Sup\'erieure\\
24 rue Lhomond, 75231 Paris Cedex 05, France\\
{\tt bilal@physique.ens.fr}\\
and\\
Institut de Physique, Universit\'e de Neuch\^atel, rue Breguet 1\\
2000 Neuch\^atel, Switzerland (address after September 1, 1998)}

\maketitle\abstracts{In these notes, I give a simple introduction to the tremendous progress that has been made
during the last few years towards the understanding of strong-coupling phenomena in quantum gauge theories and
superstring theories.}

%%%%%%%%%%%%%%%%%%%%%%%%%%%%%%%%%%%%%%%%%%%%%%%%

\section{Introduction}

In these notes, I wish to provide a simple introduction to the exciting developments of the last few years
which have made possible the understanding of strong-coupling phenomena in quantum gauge theories and
superstring theories. The key ingredients are duality and supersymmetry. Duality will be first introduced in the
classical framework of electromagnetism. However there are three immediate reasons why duality should not
work. Then I show how supersymmetry removes these objections. A key property to extrapolate from weak to
strong coupling and hence to test duality conjectures is the BPS property, which is discussed next. Then I describe
dualities in superstring theory and how they can be checked. Finally, I discuss the existence of a consistent
eleven-dimensional theory, named M-theory, that encompasses all string theories. 

\section{A first example of duality: electromagnetism}

To begin with, consider Maxwell's equations. The homogeneous equations are
\begin{equation}
\nabla\wedge B ={\partial E \over \partial t} \quad , \quad 
\nabla\wedge E =-{\partial B \over \partial t} \ .
\label{di}
\end{equation}
These two equations are perfectly symmetric under the interchange of the electric and magnetic 
fields: $E\to B,\ B\to -E$ or, more generally, defining a complex combination $(E+iB)$, under 
$E+iB\to {\rm e}^{i\alpha} (E+iB)$. In the presence of charges, however, this symmetry is broken
by   Maxwell's inhomogeneous equations:
\begin{equation}
\nabla\cdot E = \rho_e \quad , \quad \nabla \cdot B =0 
\label{dii}
\end{equation}
because we have electric charges with density $\rho_e$, but no magnetic charges.
One can restore the above symmetry, somewhat artificially at first sight,
 by introducing magnetic monopoles or magnetic charge densities $\rho_m$ and writing the second
equation as
\begin{equation}
\nabla\cdot B = \rho_m \ .
\label{diii}
\end{equation}
This innocent looking modification has important consequences. Recall that vanishing divergence
of $B$ implied the existence of a vector potential $A$ such that $B=\nabla\wedge A$, so that now
the definition of such a vector potential has become problematic. Although in classical
electrodynamics, the use of $A$ is convenient but not necessary, in quantum mechanics the
vector potential enters into the Schr\"odinger equation in a fundamental way and can be felt even in
regions of vanishing magnetic field $B$ as is witnessed by the Aharonov-Bohm effect.

This problem was solved by Dirac  long ago when he showed that even in the presence of a
magnetic monopole one can nevertheless introduce a
vector potential, although one has to admit that it has a singularity which can be taken as a string
from the monopole to infinity. This string is unobservable (by any particle of electric charge $q$) 
if and only if the magnetic charge $m$ is {\it quantized} so that $m q / ( 2\pi \hbar c)$ is an integer.
Since the elctric charges are quantized this gives
\begin{equation}
m={2\pi\hbar c \over e}\,  n_{\rm magn} \quad ,\quad q= e\,  n_{\rm el} \ .
\label{div}
\end{equation}

One might object that anyhow, neither in QED, nor in the standard model of electroweak
interactions, there are any magnetic monopoles and there is no need for
such extravagancies. However, there are many simple gauge theories, including many grand
unified theories where there are naturally elctric {\it and} magnetic charges. Actually this is the
case whenever a group without $U(1)$ factors is broken down to a group with $U(1)$'s. The
simplest example is the Georgi-Glashow model 
where $SU(2)$ is broken to $U(1)$ As shown by
't Hooft and Polyakov\cite{thooftpol}, 
there are soliton configurations in these theories that carry magnetic
charge and behave as magnetic monopoles. Indeed, solitons are non-trivial finite energy solutions
of the static field equations. As such they are extended objects with finite energy. Upon boosting
them they can also get a momentum and behave much like a  particle. Since in these models such
solitons carry a magnetic charge they are indeed magnetic monopoles. They automatically satisfy
the Dirac charge quantization condition (\ref{div}) (with $n_{\rm magn}=2$).

\section{Duality}

Suppose that in such a  theory with magnetic monopoles and electric charges we can indeed realise the duality
symmetry of exchanging every electic quantity with the corresponding magnetic quantity, then we
would in particular have a symmetry under the exchange of the minimal electric charge $e$ and
the minimal magnetic charge $m$. Using eq. (\ref{div}) with $n_{\rm magn}=2$ this gives
\begin{equation}
{4\pi\hbar c \over e} \leftrightarrow e  \ .
\label{dv}
\end{equation}
This then is an innocent looking consequence of electric-magnetic duality (if we can realise it).
We will find something  much less innocent once we realise that upon taking the square of the
previous relation we find
\begin{equation}
\alpha  \leftrightarrow {1\over \alpha} \quad {\rm where} \ \ \alpha=  {e^2\over 4\pi\hbar c}
\label{dvi}
\end{equation}
is the fine structure constant. Indeed $\alpha$ is the small expansion parameter that controls the
perturbation series of QED (and of the $SU(2)\times U(1)$ theory as well). It gives us the
strength of the electromagnetic interactions, and the fact that it is small ($\approx {1\over 137}$)
means that these interactions are relatively weak, and this is the reason why the first few orders
of perturbation theory lead to accurate results. The duality symmetry (\ref{dvi}) then is a symmetry
that exchanges weak coupling $\sim \alpha$ with very strong coupling $\sim {1\over \alpha}$.
This is precisely the aspect of duality which makes it so interesting: the hope that it may be used
to map strong coupling physics to weak coupling physics where one can do reliable computations in
perturbation theory.

Thus duality interchanges weak coupling and strong coupling, perturbative physics with
nonperturbative physics, and elementary excitations with solitons. Concerning the last point,
remember that above the magnetic monopoles appeared as solitons while the electrically charged
particles appeared as elementary (pointlike) excitations.

The question now is whether such a miraculous symmetry can really exist in a quantum field
theory.  There are a few objections that immediately come to one's mind:

1) In quantum field theory the coupling constant is not a constant but depends on the renormalisation
scale: $\alpha \equiv \alpha(\mu)$ and hence is energy dependent. It is well-known that e.g. at
LEP energies $\alpha(100 {\rm GeV}) \approx {1\over 128}$ rather than the
$\alpha(0)\approx {1\over 137}$ quoted
above. How then can one have a symmetry $\alpha \leftrightarrow {1\over
\alpha}$ if $\alpha$ is a non-trivial function of the energy scale?

2) 
In general, the quantum numbers don't match, i.e. the magnetic monopoles have spin zero, while
their hypothetical partners, the W-bosons have spin 1.

3)
A duality symmetry mapping $\alpha={1\over 137}$ to $\alpha'={1\over \alpha}=137$ is useless
since typical strong coupling physics as in QCD is at $\alpha_s\approx 1$ and mapping this to a
theory with ${1\over \alpha_s}$ is not of much help since this is still $\approx 1$.

\section{Supersymmetry}

Fortunately supersymmetry comes to our rescue concerning all three objections:

1) Supersymmetry tends to give us control over quantum, i.e. loop corrections. 
It is well known that in a supersymmetric theory certain one-loop diagrams vanish or at least 
greatly simplify because the fermionic contribution to the loop cancels the bosonic one.  
The more supersymmetry one has the more  the quantum corrections are under control. 
In particular, if one has ${\cal N}=4$ extended supersymmetry in four dimensions 
(sixteen supercharges) then there are powerful non-renormalisation theorems and the 
beta-function vanishes to all orders in perturbation theory:
\begin{equation}
\beta= \mu {{\rm d}\over {\rm d}\mu}\alpha (\mu) = 0 
\Rightarrow \alpha(\mu)=\alpha = {\rm const} \ .
\label{dvii}
\end{equation}
Then a symmetry $\alpha \leftrightarrow {1\over\alpha}$ becomes possible at all scales.

2) Supermultiplets contain different spins. Again, for ${\cal N}=4$ extended supersymmetry, 
the W-boson of spin 1 is accompanied by susy partners of spin $1/2$ and spin 0, while the magnetic 
monopole of spin 0 is accompanied by supy partners of spin $1/2$ and spin 1. Duality then simply 
exchanges the complete supermultiplets.

3) Holomorphicity \cite{seiberg}: in supersymmetric theories certain quantities depend analytically on the coupling 
constant. Actually, in a natural way a certain complex combination of the coupling $g$ and the 
$\theta$-angle
appears, namely $\tau={\theta\over 2\pi} + {4\pi i\over g^2}$.
The $\theta$-angle is the parameter that appears in the action in front of the $CP$-violating term 
$\int F_{\mu\nu}\tilde F^{\mu\nu}$. Thus it parametrizes the strength of $CP$-violation.
Certain quantities 
then depend  holomorphically on this combination $\tau$. That is to say, if such a quantity is called 
$f(\tau)$, then $f$ is a holomorphic function of $\tau$. Now the theory of holomorphic functions is 
very powerful, allowing to reconstruct the function by analytic continuation from its knowledge in 
certain domains only. Schematically what happens is the following: we might know the function $f$ 
for small couplings $g$ from perturbation theory (i.e. for values of $\tau$ with very large imaginary part). 
Duality then might allow us to obtain the function for very large $g$ (i.e. values of $\tau$ with very 
small imaginary part). Holomorphicity of $f$ then allows us to reconstruct the function $f$ for any 
value of its argument, hence for any coupling, and in particular also for couplings of the order of one. 

This program has been a tremendous success for the study of supersymmetric versions of 
QCD (${\cal N}=2$ and also ${\cal N}=1$). In particular, for the ${\cal N}=2$ susy versions \cite{SW}, 
it gives us important clues to the mechanism of quark confinement by realizing it via condensation 
of magnetic monopoles as has long been expected, as well as to chiral symmetry breaking. It has been 
possible to determine exactly the spectrum of stable particles \cite{bf} for any value of the coupling 
constant (and $\theta$-angle). Maybe rather surprisingly, it appears that there are regions 
of the parameter space (at strong coupling) where the gauge symmetry is broken, but no 
massive gauge bosons exist.  

\section{BPS states}

A favorite tool to study supersymmetric theories at strong coupling uses the so-called BPS states. 
Let me explain what they are and why they are so useful. Consider a theory in four dimensions with 
extended supersymmetry ${\cal N}\ge 2$. This means that the supersymmetry generators are not just 
a spinor $Q_\alpha$ and its conjugate $\overline Q_{\dot\alpha}$, but there are ${\cal N }$ copies of them: 
$Q_\alpha^i$ and its conjugate $\overline Q^i_{\dot\alpha}$ with $i=1, \ldots {\cal N}$. The susy algebra 
then schematically reads:
\begin{equation}
\{ Q_\alpha^i, Q_\beta^j \}  \sim  Z^{ij} \epsilon_{\alpha\beta} \ , \
\{\overline Q^i_{\dot\alpha} , \overline Q^j_{\dot\beta}
 \}  \sim  \overline Z^{ij} \epsilon_{\dot\alpha\dot\beta} \ , \
 \{ Q_\alpha^i, , \overline Q^j_{\dot\beta}
 \}  \sim  \delta^{ij} \sigma_{\alpha \dot\beta}^\mu P_\mu  . 
\label{dviia}
\end{equation}
Here $\epsilon_{\alpha\beta}$ is the antisymmetric tensor, $P_\mu$ the momentum operator.
The $Z^{ij}$ are central charges and they are combinations of the conserved abelian charges of the 
theory (i.e. electric and magnetic charges). So the anticommutation relations are of the form 
$\{Q,Q\}\sim $ abelian charges, and $\{ Q, \overline Q\}\sim$ mass. This shows that when acting on certain 
states with the appropriate (linear) relation between its mass and its abelian charges, some 
combinations of the $Q$'s and $\overline Q$'s effectively vanish. These states are then called 
BPS states. We then conclude that: 

1) BPS states satisfy  a certain linear relation between their masses and their abelian charges, 

2) BPS states are invariant under some combinations of the susy generators, meaning that 
these combinations of the generators vanish when applied on the BPS state. One says that 
a BPS state preserves part of the supersymmetries. 

3) Since some combinations of the generators vanish, one can not generate the maximal number of 
states in the supermultiplet by applying the susy generators to the state. Hence the BPS states come 
in supermultiplets that are smaller than  usual. These multiplets are called short or even supershort 
multiplets depending on how many combinations of generators vanish. While a generic susy multiplet 
has $2^{{\cal N}/2}$ states, a BPS state invariant under half the susy generators lives in a susy multiplet 
of only 
$2^{{\cal N}/4}$ states. 

It is this third property combined with the first one that make the BPS states such an ideal tool. 
Suppose we have identified a BPS state at weak coupling. It is in a short multiplet. Now continuously 
increase the coupling. Since the number of states in the susy multiplet cannot change continuously, it 
must remain constant. This means the BPS states must remain BPS, and hence still satisfy the same 
charge-mass relation characteristic for this state. This must then even be so at strong coupling, where  
otherwise it might have been very difficult to compute the mass of this state. Of course, in this type of 
argument one assumes two things: the supersymmetry is never broken, and there is no critical value of 
coupling where the BPS state suddenly might decay into other states. While this latter scenario does 
indeed happen in theories with only ${\cal N}=2$, it cannot occur for ${\cal N}\ge 4$.

\section{Dualities in superstring theory}

The so-called type II theories (IIA and IIB) have two supersymmetries in ten dimensions, which means 
32 supercharges and hence corresponds to 
${\cal N}=8$ in four dimensions, while the type I and heterotic theories have only one ten-dimensional 
supersymmetry, which means 16 supercharges and thus correspond to ${\cal N}=4$  in four dimensions.
We see that superstring theories have enough supersymmetry generators so that the above-mentioned 
decay of BPS states cannot occur. 
Thus the BPS states of superstring theory are a powerful tool to make predictions about the 
strong-coupling physics of superstring theory from the knowledge of weak-coupling physics. 
This can be used to check various duality conjectures that involve weak-strong coupling duality. 
\cite{sen}

Typically there are two types of duality: The first one is
S-duality which is strong-weak coupling duality, as was discussed for the field theories above. 
The other one is T-duality which instead changes some other parameters of the theory like 
compactification radii. 
While S-duality is non-perturbative in nature (it exchanges perturbative and non-perturbative physics), 
T-duality can be checked order by order in perturbation theory. 

The simplest example for T-duality is superstring theory compactified on a circle of radius $R$. 
Then T-duality maps this to a superstring theory compactified on a circle of radius ${\alpha'\over R}$. 
Here $\alpha'$ is the parameter that sets the length scale for the string  
$\alpha' \approx (10^{-32} {\rm cm} )^2$.
The masses of the string states are ${\cal M}^2_{\rm string} \sim \left({n\over R}\right)^2
+ \left({m R\over \alpha'}\right)^2$ plus contributions from the oscillator terms. The first term is due to the momentum
modes around the circle, while the second term comes from configurations where the string winds $m$ times around the
circle. Let me explain this a little more. If a  coordinate, say $X^9$ is compactified on a circle of radius $R$, one has to identify $X^9$ and $X^9 +
2\pi R$. A plane wave ${\rm e}^{i p_9 X^9}$ than is single-valued provided the momenta are quantized as $p_9={n\over R}$.
This explains the first term in the mass formula. A string that winds $m$ times around the circle has a minimal length
$2\pi mR$, and this contributes to the energy an amount $2\pi m R \times {1\over 2\pi \alpha'}$ where ${1\over 2\pi
\alpha'}$ is the tension of the string. This explains the second term in the mass formula. There are additional terms giving
the contributions from the harmonic oscillator modes, but these are independent of the compactification radius.
One clearly sees that $R \leftrightarrow {\alpha'\over R}$ is a symmetry of the mass formula provided one also
exchanges the quantum numbers $n$ and $m$, i.e. one exchanges the momentum and winding modes. In particular this
means that perturbative string theory cannot distinguish between a compactification radius $R$ and ${\alpha'\over R}$.
In a way this means that there is a minimum length one can explore, namely $\sqrt{\alpha'}$.

As we saw, it is very easy to get evidence for T-duality. How can one find similar evidence for the non-perturbative
S-dualities? One way is to look at the low-energy effective actions of the string theories that are supposed to be
related by this duality. The low-energy effective action describes the effective quantum field theory of the massless
modes of the string theory. Let us consider an example. On the one hand, consider type I superstring theory 
which has a gauge group ${\rm SO}(32)$ and
massless fields that are vector fields in the adjoint representation of this group, a symmetric traceless tensor (graviton),
an antisymmetric tensor and a scalar (dilaton). The effective action describing the dynamics of these massless fields at
low energies is constrained by supersymmetry to be ${\cal N}=1$ supergravity in ten dimensions coupled to ${\rm
SO}(32)$  super Yang-Mills theory. On the other hand consider the heterotic string theory with gauge group 
${\rm SO}(32)$ . It has the same massless fields as the type I theory, and the same supersymmetry consideration also
constrain the low-energy effective action to be ten-dimensional supergravity coupled to ${\rm SO}(32)$ super
Yang-Mills. There is however some freedom in the way the dilaton $\phi$  appears in these actions.
Looking at the actions one finds that 
$S_{\rm eff}^{\rm I}(\phi^{\rm I}, \ldots) = S_{\rm eff}^{\rm het}(\phi^{\rm het}, \ldots)$ provided one identifies 
$\phi^{\rm I}= - \phi^{\rm het}$. But the string coupling constant is given by the exponential of the dilaton, so that one
finds
$g_s^{\rm I} = {\rm e}^{\phi^{\rm I}} =  {\rm e}^{-\phi^{\rm het}} = {1\over  g_s^{\rm het} }$.
This then prompts the conjecture: Strongly coupled type I theory is equivalent to weakly coupled heterotic ${\rm
SO}(32)$ theory with $g_s^{\rm I}={1\over  g_s^{\rm het} }$.

\section{Testing S-duality in superstring theory}

How can we further test such an S-duality conjecture? We need to know something about the strong-coupling spectrum
of the ${\rm SO}(32)$ type I theory and compare it with the weak-coupling spectrum of the ${\rm SO}(32)$ heterotic
theory. To do this we will use BPS states.  Let me first describe which BPS states we are going to use.

In addition to the fundamental strings there exist also solitonic states, similar to domain walls, which are hypersurfaces
in space (-time) on which open strings can end (see Fig. 1). Because the string end-points are restricted 
to these hyperplanes, the
strings have Dirichlet boundary conditions in the directions normal to the hyperplane (while they still have the ordinary
Neumann conditions in the directions tangential to the hyperplane). This is the reason why these hyperplanes are called
Dirichlet branes, or D-branes for short. If they have $p$ space-dimensions they are called D$p$ branes.

\begin{figure}[htb]
\vspace{9pt}
\centerline{ \fig{4.5cm}{4.5cm}{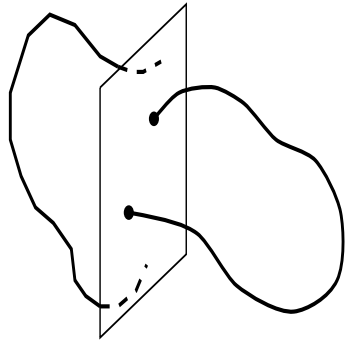} }
\caption{A D$p$-brane is a hypersurface with $p$ space-like and
one time-like dimensions on which open strings with Dirichlet
boundary conditions end.}
\label{singleD}
\end{figure}

The fact which interests us most here is that such D-branes break half of the supersymmetries and preserve the other
half. Hence they are BPS states. A D-brane of infinite extent has an infinite energy/mass, but  if one compactifies $k$
dimensions, say on a $k$-dimsional torus, then a D$p$ brane with $p=k$ can wrap around this torus and have a stable
configuration of finite area and hence of finite energy/mass. Also, for $p=0$,  one has D0 branes that always have finite
mass. Thus we see that D-branes may behave like states of finite energy just like particles or solitons. 

\begin{figure}[htb]
\vspace{9pt}
\centerline{ \fig{8.0cm}{5.5cm}{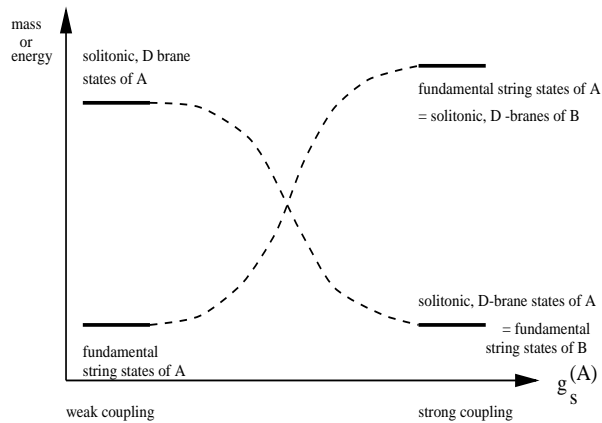} }
\caption{As the coupling is changed from weak to strong one may follow the BPS states. Typically heavy states become
light and vice versa.}
\label{bpsinterpolation}
\end{figure}

We may now look
at the spectrum of finite energy states in  one theory, call it theory A. Typically at weak coupling  ($g_s\ll 1$)
one will have relatively
``light" states which are the (perturbative) fundamental string multiplets (with masses of the order of
$m_s=1/\sqrt{\alpha'}$), as well as very heavy states which are the solitonic D-branes (with masses of the order of
$m_s/g_s$). Now imagine one increases the coupling constant $g_s$. For an arbitray state one does not know what will
happen, but for a BPS state, as explained above, it will continue to exist with the same multiplicity and the same mass to
charge relation (which for D-branes contains the inverse power of the coupling constant). Typically, as we go to strong coupling the
formerly heavy states  become relatively light while the formerly light states  become heavy, see Fig. 2. This now looks
very similar to the original weak coupling spectrum except that perturbative states and non-perturbative states have
changed their places. We may now compare this spectrum of theory A at strong coupling with the weak-coupling BPS
spectrum of some other theory B. If these spectra match with the correct multiplicities, this constitutes a very strong
indication that theory A at strong  coupling is equivalent (dual) to theory B at weak coupling, i.e. for S-duality between
theories A and B.

There exist also other dualities that mix S and T, i.e. at the same time one exchanges strong and weak coupling one also
has to exchange some of the dimensionful parameters, like $R\to {\alpha'\over R}$. These go under the name of 
U-dualities.

\section{Even stranger : M-theory}

Consider now type IIA superstring theory. This is a closed string theory with 32 supercharges, and its low-energy
effective action is IIA supergravity in ten dimensions. It has been known for long that this ten-dimensional IIA
supergravity can be obtained by dimensional reduction from eleven-dimensional supergravity. This is purely within the
context of field theory. In the string context, this relation however remained mysterious for a long time. Although type II
theories are theories of closed strings, it makes sense to also introduce D-branes on which open strings can start and
end. This is particularly motivated by the fact that a D$p$ brane carries a charge with respect to the
antisymmetric $p+1$ index tensor field that
may appear in the Ramond-Ramond sector of the string theory. In type IIA $p=0,2,4,6,8$ occurs, 
so we need D0, D2, D4, D6 and D8
branes. 

In particular in type IIA superstring theory, there are D0 branes which actually behave as point particles and are also 
called D particles. Their
mass is given by $m_{D0}={m_s\over g_s}$ where $m_s=1/\sqrt{\alpha'}$ is the string mass and $g_s$ the string coupling
constant. Several such D0 branes may form a bound state with zero binding energy. Such bound states at threshold are
possible due to the supersymmetry. Actually although not completely proven, we now have very good evidence that
there is exactly one such threshold bound state for any number $n$ of D0 branes. Such a bound state then has a mass
$n{m_s\over g_s}$. We see that the mass spectrum of the D0 branes  consists of equidistant levels 
at $n{m_s\over g_s}$ with $n=1, 2, \ldots$. Such an equidistant spectrum  is typical for a Kaluza-Klein spectrum. Indeed,
if one has a theory in $d+1$ dimensions with a massless state, and one compactifies one dimension on a circle of radius
$R$ than in $d$ dimensions one sees an infinite tower of massive states with masses $n {1\over R}$. It is now very
tempting to speculate that the D0 brane spectrum of the IIA superstring indeed is a Kaluza Klein spectrum that originates
from eleven dimensions and that the compactification radius is $R\equiv R_{11}={g_s\over m_s}$. Then as one takes the
string coupling $g_s$ to infinity, $R_{11}\to \infty$ and one recovers a continuus spectrum and the uncompactified
eleven-dimensional theory. 

The natural candidate
in eleven dimension is the eleven-dimensional supergravity. If this is the right guess then all Kaluza Klein states must
come as BPS multiplets of 256 states because they would arise from the elevendimensional supergraviton which fills ut
a 256 dimensional supermultiplet. But the D0 branes and their bound states are BPS and precisely come in multiplets of
256 states ! Thus we have found that type IIA string theory, in the strong coupling limit, and at low energies, is
eleven-dimensional supergravity. Of course, at arbitrary energy scales, type IIA theory has more states than just D0
branes.
It is also known that eleven-simensional supergravity is not likely to be a consistent
quantum theory at all energy scales. On the other hand, type IIA string theory is. So this led to the conjecture that there
is some as yet mysterious theory, named M-theory, that is a consistent quantum theory at all energy scales and whose
low-energy limit is eleven-dimensional supergravity. More precisely, the preceeding considerations lead to the following
identification:
M-theory compactified on a circle of radius $R_{11}$ is identical to type IIA superstring theory with coupling constant
$g_s=R_{11}/\sqrt{\alpha'}$.
The non-trivial conjecture is that this M-theory, if not compactified, has eleven-dimensional Lorenz invariance, just as
has its low-energy limit, the eleven-dimensional supergravity.

The conjecture of the existence of such an eleven-dimensional M-theory is very useful to derive many dualities in
superstring theory. For example a T-duality in M-theory can exchange $R_{11}$ and $1/R_{11}$. But in IIA superstring
theory this corresponds to exchanging $g_s$ and $1/g_s$ which is S-duality, etc. One can derive a whole web of
dualities of string theories in ten dimensions and even more for string theories compactified on certain manifolds. For
example type IIA compactified on the four-dimensional surface named $K3$ is a six dimensional theory where half of the
initial 32 supersymmetries have been broken by the compactification so that only 16 remain. On the other hand, the
heterotic theory compactified on a four-dimensional torus (four dimensions compactified each on a circle) still has all
sixteen supercharges that one had initially in ten dimensions. So the two six-dimensional theories both have sixteen
supercharges, and one indeed shows that they are dual to each other.

\section{Conclusions}

We have seen that weak-strong coupling duality is realised in certain quantum field theories. 
Even more striking, it plays a
central role in the large web of dualities in superstring theory. This allows us to often get complete
non-perturbative information about the strong-coupling physics. In all of these developments,
supersymmetry plays a crucial role in establishing
the duality properties. Even after susy breaking we might still control many strong-coupling phenomena.
And if at worst, nature turned out not to be susy, we have at least learned many important lessons
about the possible strong-coupling behaviour of quantum field theories.

\section*{Acknowledgments}
It is a pleasure to thank the organizers of the Moriond meeting and in particular  Tran Thanh Van
and Pierre Fayet 
for the invitation to give this talk. This work was
partially supported by TMR contract FMRX-CT96-0090.

\end{document}